\newcommand{\CLASSINPUTtoptextmargin}{2cm}
\newcommand{\CLASSINPUTbottomtextmargin}{3cm}
\documentclass[journal, a4paper]{IEEEtran}

\usepackage{cite}

\usepackage{graphicx}
\DeclareGraphicsExtensions{.eps}
\usepackage{setspace}
\providecommand{\PSforPDF}[1]{#1}
\usepackage[cmex10]{amsmath}
\usepackage{amsfonts,amssymb,amsthm}
\usepackage{times}
\usepackage{color}

\PSforPDF{\usepackage{psfrag}}
\usepackage{algorithmic}
\usepackage{algorithm}
\usepackage{epstopdf}

\usepackage{array}
\usepackage{mdwmath}
\usepackage{mdwtab}
\usepackage{eqparbox}
\usepackage{nicefrac}
\usepackage{exscale,relsize}
\usepackage{subfig}
\usepackage{url}

\usepackage{fancyhdr}
\begin{document}
\title{Coordination protocol for inter-operator spectrum
  sharing based on spectrum usage favors}
\author{\IEEEauthorblockN{Bikramjit Singh, Konstantinos Koufos and
    Olav Tirkkonen} \\ %  <-this % stops a space
\IEEEauthorblockA{Communications and Networking Department, Aalto
University, Finland} \\ 
Email: \{bikramjit.singh, konstantinos.koufos,
    olav.tirkkonen\}@aalto.fi}

\maketitle
\thispagestyle{plain}
\fancypagestyle{plain}{
\fancyhf{}
\fancyfoot[C]{}
\fancyfoot[R]{}
\renewcommand{\headrulewidth}{0pt}
\renewcommand{\footrulewidth}{0pt}
}
\pagestyle{fancy}{
\fancyhf{}
\renewcommand{\headrulewidth}{0pt}
\renewcommand{\footrulewidth}{0pt}

\begin{abstract}
Currently, mobile network operators are allocated spectrum bands
on an exclusive basis. While this approach facilitates interference
control, it may also result in low spectrum utilization efficiency. 
Inter-operator spectrum sharing is a potential
method to enhance spectrum utilization. In order to realize it, a
protocol to coordinate the actions of operators is needed. We propose
a spectrum sharing protocol which is distributed in 
nature, it does not require operator-specific information exchange and
it incurs minimal communication overhead between the operators. 
Operators are still free to decide whether they share spectrum or not
as the protocol is based on the book keeping of spectrum usage favors, 
asked and received by the operators. We show that operators can enhance their
QoS in comparison with traditional orthogonal spectrum allocation
while also maintaining reciprocity i.e. no operator benefits over the
other in the long run. We demonstrate the usability of the proposed
protocol in an indoor deployment scenario with frequent network load
variations as expected to have in small cell deployments.
\end{abstract}

\section{Introduction}
\label{sec:Introduction}
To meet the increasing data traffic demand in a timely manner, a
viable solution is a shared use of radio frequency (RF) spectrum where
multiple independent users can utilize the same RF resources provided
that they do not generate destructive interference to each
other~\cite{EC2012}. Shared spectrum use between mobile network
operators has significant business potential particularly in small
cell deployments. For the time being, spectrum is exclusively
assigned to the operators. In a  more flexible regulatory world, it is
envisioned that operators may agree to let other operators use their
spectrum. For instance, different operators can provide high speed
data access at different parts of shopping  malls. In that case,
inter-operator interference is low and operators can entertain
the benefit of higher available bandwidth without disturbing each other.

A coordination protocol is a distributed method to enable spectrum
sharing between peer networks~\cite{Tim2013}. It requires a logical
connection between the different networks e.g. over-the-air, via the
core network, etc. Existing coordination protocols for inter-operator
spectrum sharing assume either operator-specific information exchange or that
operators agree beforehand on some spectrum allocation which is 
maintained under the threat of punishment. These 
attributes can be problematic because on the one hand, operators are
competing entities and on the other, under frequent network load
variations static spectrum allocation has poor performance. 

Operators are expected to share spectrum for a long 
time. Due to the fact that an operator has a persistent and publicly
known identity, the operators can learn from each other's behavior. 

We propose a coordination protocol that realizes spectrum sharing by 
means of a virtual monetary economy~\cite{Engineer} based on a
rudimentary currency in 
terms of spectrum usage favors, asked and received by the
operators. In this perspective, operators are free to decide whether
they take part into the spectrum usage negotiations or not. According to the
proposed protocol, operators with low load fulfil spectrum usage
favors to heavily-loaded operators. Operators are needed to maintain the reciprocity, 
and the operators with granted favors in the past will return these
favors in future. In this way, all
operators offer better QoS in comparison with static spectrum
allocation without revealing their specific performance indicators nor
making any agreement beforehand. 
%The proposed coordination protocol does not require a central entity
%in charge (e.g. a spectrum broker) to handle the interactions between the operators.

\section{Proposed coordination protocol}
Let us assume that an operator can construct a number that describes the
QoS offered to its users. This kind of number is usually 
referred to as network utility. The network utility function can, for
instance, be defined as a linear combination of average and cell edge
performance. Note that the proposed coordination protocol does not
require that the operators employ the same utility function nor that
the operators are aware of each other's utility function.
 
Before asking/granting a favor, an operator has to evaluate the effect
the opponent operator has on its 
utility. In order to do that an operator should measure the amount of interference it
receives from the opponent. For downlink transmissions, this
functionality can be a simple extension of LTE handover
measurements. For example, the operator may ask its users to measure
on the interfering signal levels and report them to the serving base
station. Note that this kind of functionality does not require any
signaling between serving and interfering base stations, as in
the regular handover procedures. 

For different spectrum sharing scenarios there should be different types of
spectrum usage favors asked and taken among the operators. According
to the limited spectrum pool scenario, there is a shared pool of resources 
available to use by a certain number of users~\cite{D51}. With
limited spectrum pool we view a single type of spectrum usage favor:
\begin{itemize}
\item Operator asks the opponent operator for permissions to start using a resource
  from the pool on an exclusive basis. 
\end{itemize}

In a mutual renting scenario, each operator owns exclusively a certain
amount of resources but there can be mutual agreements between
operators for resource utilization. In that case we view two different
kind of spectrum usage favors: 
\begin{itemize}
\item Operator asks the opponent operator for permission to start
using one of the opponent's resources. 
\item Operator asks the opponent operator for permission to start
  using one of the opponent's resources on an exclusive basis. 
\end{itemize} 

In general, the spectrum usage favor should be granted for a certain
time interval that has to be agreed among the different
entities. During that time interval, the 
operators do not renegotiate the usage of that particular
resource. After that interval, the resource allocation falls back to  
the state it had before granting the favor, i.e. in limited spectrum
pool both entities should utilize the resource and in mutual renting the
entity granting the favor can start using the resource on an
exclusive basis.

Operators are self-interested entities. They will ask/grant a favor only if, in the long run, they expect to get more benefits than losses. To do so, one way is to investigate the history of previous interactions with the opponent  operator and accordingly take the actions. For example, before asking for a favor, the operator checks whether its immediate
utility gain is larger than its average utility loss 
over the history of previous interactions. 
In a similar manner, an operator grants a favor only if its immediate
utility loss is smaller than its average past utility gain.
In order to maintain reciprocity, operators should grant about the
same number of favors over the time. For that purpose, we define a
positive integer number $S$ that determines the maximum allowable
number of outstanding favors. 

Overall, the proposed decision making process avoids immediate 
punishment and it is also forgiving, resembling an extended time
period tit-for-tat 
retaliation strategy~\cite{Axelrod1981}. As soon as the opponent
starts granting favors again, the 
operator would also be willing to cooperate and grant favors
back. 

% One should keep in mind a possible scenario where an
% operator takes $S$ more favors than the opponent and then stops
% to grant any favor. One way to prevent ever-lasting benefits due
% to cheating is to enforce a fallback into the
% orginal spectrum allocation after a certain number of timeslots
% i.e. traditional orthogonal allocation under mutual renting scenario
% and full spread under limited spectrum pool 
% scenario. Coordination protocols incorporating dishonest behaviors are
% left for future study.

Before illustrating the usability of the proposed protocol, we note
few more aspects which are not treated in this work. In general,
spectrum usage favors can be of different priorities. For example,
when an operator cannot meet its QoS, it may add a high priority in
the favor it asks for and subsequently, provided that the favor is
granted, it will also get a high penalty in the future due to the priortized favors it has received in the past. 
Also, operators can ask/grant favors for a
variable time interval which should be optimized taking into account
the history of previous interactions and the traffic model. We leave
such kind of protocol features for later study. In this poster, we
consider spectrum usage favors of equal priority which are valid for
a fixed time interval. 

\begin{figure}[!t]
  \centering
  \includegraphics[width=0.44\textwidth]{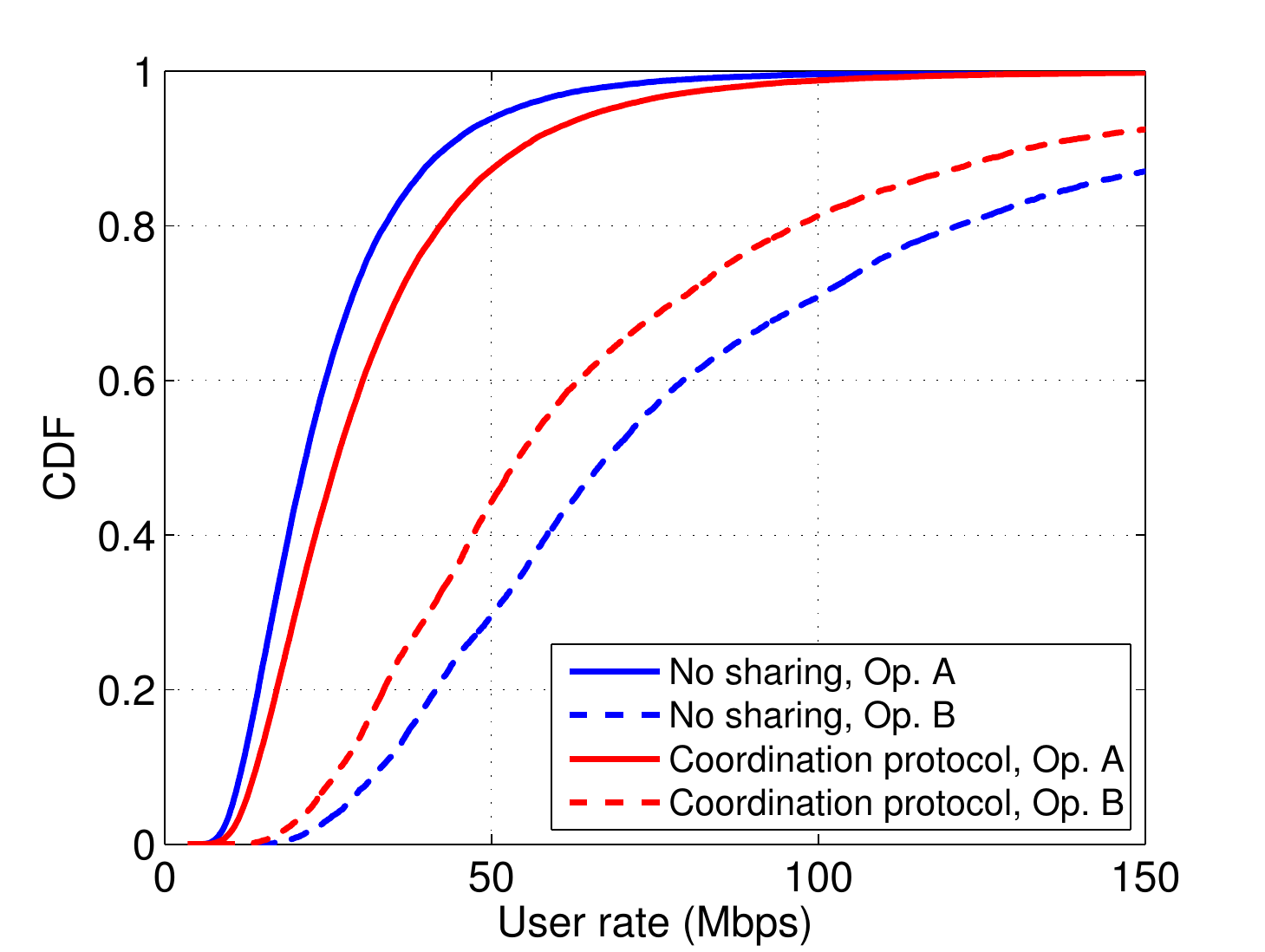}
  \caption{Rate distribution for Operator A with higher mean load and Operator B with lower mean load using orthogonal spectrum sharing and the coordination protocol in a limited spectrum pool scenario.}    
  \label{fig:figure1}
\end{figure}

\section{Results}
We consider a deployment scenario with two 
operators, Operator A and Operator B coexisting in spectrum and
offering high speed data services at different parts of a
single-story building. Each operator is allocated one component
carrier to ensure coverage while there are also six component carriers
belonging to a shared spectrum pool. We consider a scenario with high
inter-operator interference and load asymmetry. 
We evaluate the performance of the proposed coordination protocol over
a finite time horizon with $1000$ different deployment snapshots. Rest of the parameter settings can
be found in~\cite{D51}.

In Fig.~\ref{fig:figure1} the rate distributions for the users of Operator A and Operator B are depicted, when Operator A has a higher mean load than Operator B. In this scenario, Operator A receives more favors than Operator B. In Fig.~\ref{fig:figure2}, the rate distributions for the users of Operator A are depicted over the full simulation time, where the operator experiences both high and low load states. Operator A has received more component carriers, when it has high load during the course of simulation. Overall, better QoS may be offered when using the coordination protocol, than with static orthogonal sharing. The user rate distribution curves for Operator B follow the same trend.

\begin{figure}[!t]
  \centering
  \includegraphics[width=0.44\textwidth]{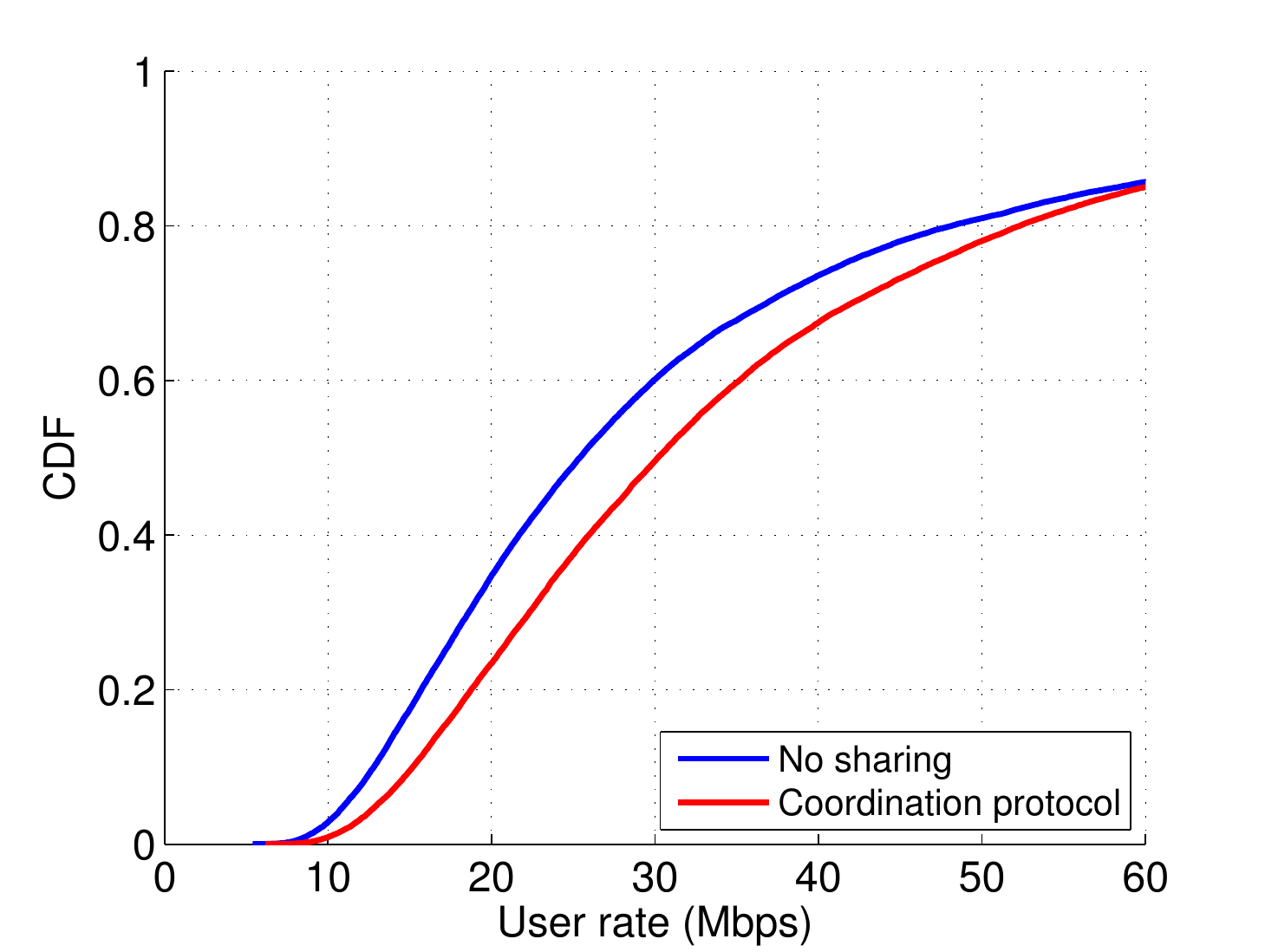}
  \caption{Rate distribution for Operator A with load variations using orthogonal spectrum sharing and the coordination protocol in a limited spectrum pool scenario.}    
  \label{fig:figure2}
\end{figure}

\section*{Acknowledgment}
This work has been performed in the framework of the FP7 project ICT
317669 METIS, which is partly funded by the European Union.

\end{document}